# Omnidirectionally manipulated skyrmions in an orientationally chiral system

Jiahao Chen [a, †], Wentao Tang [b, †], Xingzhou Tang [a, †] Yang Ding [a], Jie Ni [a], Yuxi Chen [d], Bingxiang Li [a, c, *], Rui Zhang [b, *], Juan de Pablo [d, e], Yanqing Lu [c, *]

[a] College of Electronic and Optical Engineering & College of Flexible Electronics (Future Technology), Nanjing University of Posts and Telecommunications, 210023 Nanjing, China

[b] Department of Physics, The Hong Kong University of Science and Technology, Clear Water Bay, Kowloon, 99999 Hong Kong, China;

[c] National Laboratory of Solid State Microstructures, College of Engineering and Applied Sciences, Nanjing University, 210093 Nanjing, China

[d] Pritzker School of Molecular Engineering, University of Chicago, 60637 Illinois, USA

[e] Center for Molecular Engineering, Argonne National Laboratory, 60439 Illinois, USA

† These authors contributed equally to this work

*Corresponding author: bxli@njupt.edu.cn; ruizhang@ust.hk; yqlu@nju.edu.cn

**Abstract**

Skyrmions, originally from condensed matter physics, have been widely explored in various physical systems, including soft matter. A crucial challenge in manipulating topological solitary waves like skyrmions is controlling their flow on demand. Here, we control the arbitrary moving direction of skyrmions in a chiral liquid crystal system by adjusting the bias of the applied alternate current electric field. Specifically, the velocity, including both moving direction and speed can be continuously changed. The motion control of skyrmions originates from the symmetry breaking of the topological structure induced by flexoelectric-polarization effect. The omnidirectional control of topological solitons opens new avenues in light-steering and racetrack memories.

# Introduction

Solitons possess the remarkable ability to maintain their waveform and velocity intact, even after colliding with one another[1], which have found extensive applications including plasma physics[2], high-energy electromagnetics[3], and nonlinear optics[4]. Nematic liquid crystal (NLC) is a nonpolar fluid with a long-range orientational order. The reconfigurable director field $\hat{n}$, called "director", describes the average local alignment direction of rod-like liquid crystal (LC) molecules, which enables the formation of solitary waves of molecular reorientation[5]. Anisotropic dielectric properties of NLCs, namely, dielectric anisotropy $\Delta\varepsilon = \varepsilon_\parallel - \varepsilon_\perp$, with $\varepsilon_\parallel$ measured along and $\varepsilon_\perp$ perpendicular to the optic axis, also provide a testbed to study electrically generated solitons[5]. Introduction of chirality leads to chiral nematic liquid-crystal (CNLC) in which the director aligns in a helical pattern around certain axis. The distance over which $\hat{n}$ rotates around helically by $2\pi$ is known as the pitch, denoted by $p$[6-13]. Chirality enriches the topological behavior of solitons in LCs, enabling a transition from non-topological to topological states and giving rise to diverse types of topological solitons[14-18]. Skyrmions, originally proposed by Tony Skyrme, are particle-like entities associated with baryons, including combinations of heavy baryons and resonances[19,20]. They are common topological solitons in CNLCs.

The investigation of motion of both non-topological and topological solitons has been a major research focus. The non-topological soliton, "directrons", based on flexoelectric effect[5] or electrohydrodynamic instability[21], can move directly with a high speed. They can also achieve a 90° turning[22] or a range of the motion angles within 50°[21] by tuning the electric field. For topological solitons, like Skyrmions, are typically static in the early stages of research[23]. Motion of topological solitons was first achieved by applying modulated electric field[15,16]. By giving a pre-alignment at the boundaries of the confined CNLC film, it is possible to achieve the motion of skyrmions along the direction perpendicular to the orientation[24]. Other approach to control topological soliton motion involves using optical manipulation[17,18]. However, omnidirectional motion of solitons is still a grand challenge.

In this work, we present the first realization of 360°- tunable motion direction of skyrmions in a CNLC system. Specifically, by vertically applying a square alternating current (AC) electric field with a bias voltage to a confined CNLC film, the topological solitons within the cell can move in any direction and stay at any fixed point. In this way, the trajectories can be predesigned depending on the relationship between motion directions and the applied bias voltage. The motion speed, direction, and size of the skyrmions are all influenced by the amplitude and direction of the bias voltage. The structure and moving direction of these solitons are analyzed and explained by numerical simulations, which demonstrate that an additional polarization field $\boldsymbol{P}$ emerging in the system exhibits a linear coupling with the electric field $\boldsymbol{E}$. The spatial polarization arises from the strong flexoelectricity effect of the material, which can alter the velocity components of the skyrmions in different directions, thereby changing their direction of motion. To the best of our knowledge, omnidirectional motion control of solitons in a two-dimensional plane through external fields have not yet been achieved in any other system.

## Results and discussions

### Materials and experimental design

Skyrmions are generated and propagate in the chiral nematic liquid crystal when a square-wave electric field is applied. Here, we utilize a chiral nematic liquid crystal composed of a nematic liquid crystal host (DP002-026, 97 wt%) and a chiral dopant (S811, 3 wt%) with an equilibrium helicoidal pitch $p \approx 2.7\,\mu\text{m}$, as measured using a stripe–wedge Grandjean–Cano cell (supplementary Fig. S7). The CNLC is injected into the liquid crystal cell, which consists of two glass substrates, in its isotropic phase. As depicted in Fig. 1a, the substrates are coated with indium-tin-oxide (ITO) to serve as transparent electrodes. Alignment polyimide is spin-coated and rubbed onto the substrates to provide planar anchoring, ensuring the helical axis of the CNLC at the LC/glass interface aligned along the $z$-axis. The cell gap is controlled using spherical silica particles with a diameter of $d = 7.0 \pm 0.5\,\mu\text{m}$. An alternating current (AC) square voltage $U_0 = 0 \sim 10\,\text{V}$ with a bias voltage ($U_\text{b}$) ranging from $-11 \sim 11\,\text{V}$ is applied to the cell at a frequency of $f = 20\,\text{Hz}$, Fig. 1b, at which the nematic host exhibited a positive dielectric anisotropy ($\Delta\varepsilon > 0$). The electric field is oriented normal to the glass substrate and along the $z$-axis, $\boldsymbol{E} = (0, 0, E)$.

**Generation of topological solitons**

For simplicity, we use AC square voltages in the absence of bias at $T = 34°C$ to generate the skyrmions. Due to the rubbing substrate with the planar anchoring, without an external field, a uniform state forms in the cell, Fig. S1 b (I). When a voltage $U_0$ below the threshold voltage, $U_{th}$, $U_0 < U_{th}$ is applied, the texture of the CNLC film remains uniform, namely the periodic helical structure still exists, for $f = 20\,\text{Hz}$, $U_{th} = 5.6\,\text{V}$, Fig. S1 b (I). When gradually increasing the amplitude of the electric field, for positive dielectric liquid crystal, the director field starts to tilt out-of-plane and a roll pattern appears at $5.6\,\text{V} < U_0 < 7.6\,\text{V}$, Fig. S1 b (II); as the voltage grows, the CNLC film exhibits a fingerprint texture, Fig. S1 b (III); with further increase in $U_0$, cholesteric fingerprints gradually shrink and eventually skyrmions appear, Fig. S1 b (IV), and Supplementary Movie 1. Specifically, some cholesteric fingers shrink and form closed loops, which exhibit larger sizes compared to the skyrmions. These loops can also transform into skyrmions by increasing the electric field, supplementary Fig. S2 and Supplementary Movie 2. Gradually increasing the voltage results in few skyrmions being generated. Therefore, in the experiments, we typically induce a large number of skyrmions by abruptly changing $U_0$ from 0 to 10 V, Supplementary Movie 3. At a higher voltage, the topological structures of the skyrmions are destroyed and they eventually disappear, Fig. S1 b (V), which means that the director of the CNLC is forced to align normally to the substrates. When the frequency of the electric field is less than $10^4\,\text{Hz}$, the threshold voltage of each phase remains approximately constant as the frequency increases. When the frequency reaches $10^5\,\text{Hz}$, the threshold voltage of each phase sharply increases, Fig. S1 a.

Through numerical simulations, we examined the structure of skyrmions when an AC square voltage is applied in the absence of bias voltage, Fig. 1c. External fields and confinements introduce frustration to the system, leading to significant structural changes in the LC orientational field. Fig. 1c illustrates the cross-section of the director field of the skyrmion along the *x-z* plane and *x-y* plane when the voltage reaches its peak within one cycle. At the middle plane of the cell, the effect of surface anchoring is small. Hence, the directors align vertically at the perimeter of the skyrmion and undergo a $\pi$ rotation with respect to the skyrmion center. The helicity of the skyrmion at the center is equal to $\pi/2$ and the deformation is mainly double twist without splay deformations in the *xy*-plane. However, when close to the substrate, the planar boundary forces the director to lay down and draw close to the *x* direction. Extra splay deformations are induced in the skyrmion, and the helicity of the skyrmion changes as well. To be compatible with the chirality, the directors exhibit twist deformations along the thickness direction which results in a continuous variation of the helicity. Specifically, in Fig. 1c, the director field near the top substate exhibits a helicity larger than $\pi/2$, and the helicity gradually decreases and becomes less than $\pi/2$ near the bottom substrate. The POM images from both experiment and simulation are depicted in Fig. 1d, which show a good agreement.

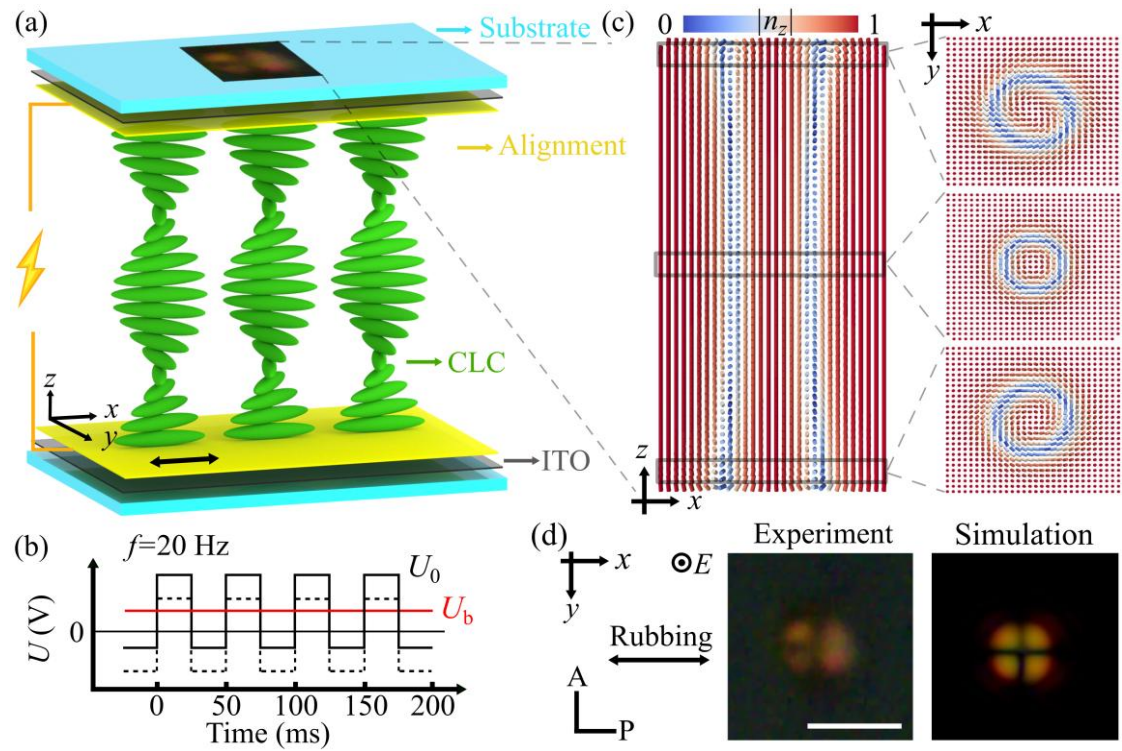

**Fig. 1 Schematic of CNLC cell and field-induced skyrmions. a**, Experimental setup for generating and observing skyrmions. The black double-headed arrow represents the direction of rubbing; **b,** Waveform of the applied electric field; **c,** Simulation result of an individual skyrmion, showing asymmetry along the *z*-axis; **d,** Comparison of POM images of a single skyrmion from experiment and simulation. Scale bar 5 μm.

**Electrically controllable velocity of skyrmions and corresponding explanation.**
The direction of the skyrmion propagation can be controlled by adjusting a bias voltage, while the speed can be adjusted by changing the frequency. In this section, we manipulate the bias voltage and frequency and observe the changes in the propagation direction and speed, as shown in Fig. 2a and Supplementary Movie 4. When the frequency is 20 Hz, we utilize a square-wave AC field with an amplitude of $U_0 = 10\,\text{V}$. When $U_b = 0$, the skyrmion moves along the $y$ direction, and it moves along the $-y$ direction when the bias voltage is set to $\pm 11\,\text{V}$. When $U_b > 0$, the direction of propagation is between 0 and $\pi$, whereas a negative $U_b$ changes the direction between 0 and $-\pi$. Note that increasing $U_b$ results in a counterclockwise orientation change, while decreasing $U_b$ results in a clockwise turn. Fig. 2a shows the direction of motion under different bias voltages $U_b$ with a fixed frequency $f = 20\,\text{Hz}$. We define the direction of motion when $U_b = 0$ as the original direction, $\alpha=0$. Both experiment and simulation show that the bias voltage ranging from $-11\,\text{V}$ to $11\,\text{V}$ results in a change in $\alpha$ from $-\pi$ to $\pi$, Fig. 2b. We do not show higher bias voltages because the structure of the skyrmions become unstable and will gradually elongate when $|U_b| > 11\,\text{V}$.

The speed, as shown in Fig. 2c, is at maximum ($3.1\,\mu\text{m/s}$) when the bias voltage is zero and decreases as the absolute value of the bias voltage increases. We believe that this occurs because a large bias voltage leads to a stronger rotation of the skyrmion, consuming more energy to counteract the elastic forces, thereby decreasing the kinetic energy. The change in frequency does not influence the propagation direction but does affect the speed. Fig. 2d and Supplementary Movie 5 show that the speed decreases as the frequency increases. When the frequency is $10\,\text{Hz}$, the speed is $4.2\,\mu\text{m/s}$. It reduces to $0.5\,\mu\text{m/s}$ when the frequency is $50\,\text{Hz}$. The response time of the nematic host is on the order of tens of milliseconds and cannot respond to frequency with kHz level, thus resulting in a lower speed. Thus, by combining the manipulation of bias voltage and frequency, we can effectively control the velocity of the skyrmions.

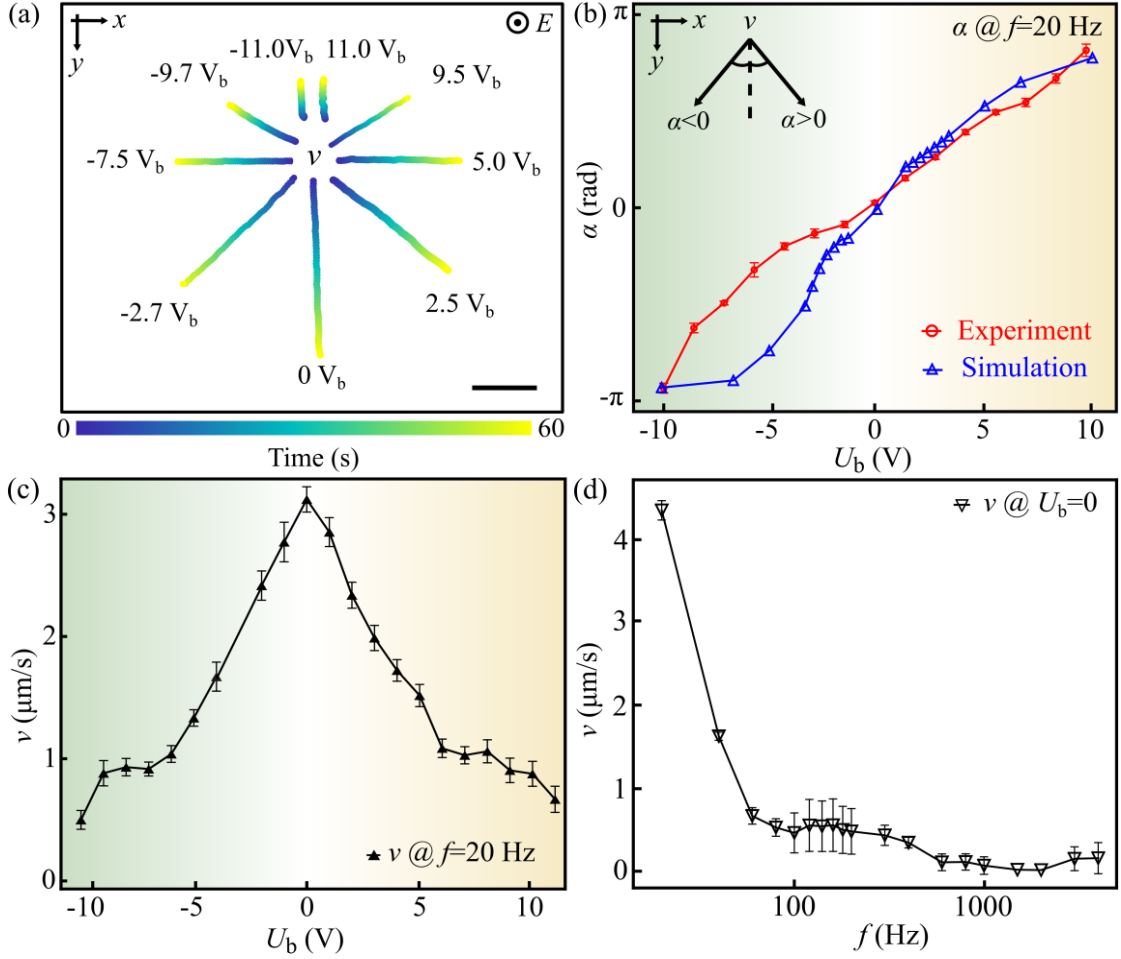

**Fig. 2 Velocity of skyrmions under different electric conditions. a,** 360° control of emission direction by adjusting the bias voltage; **b,** The direction of skyrmion motion changes with the bias voltage at a frequency of 20 Hz; **c,** The speed of skyrmions under different bias voltages; **d,** The speed of skyrmions influenced by frequency.

To eliminate the influence of backflow effects on the direction of skyrmions motion, we first calculate the Ericksen number $Er=\dfrac{\eta v L}{K}$ of the system. The rotational viscosity of the material is $51\times10^{-3}$ Pa·s [25] and the average elastic constant $\bar{K} = \dfrac{K_{11} + K_{22} + K_{33}}{3}$ is approximately $17\times10^{-12}$ N. For the velocity and characteristic length, we use the value of the average velocity and the diameter of the skyrmions, $v \approx 2\times10^{-6}$ ms$^{-1}$ and $L \approx 3\times10^{-6}$ m. Taking these into account, The Ericksen number is $\sim 0.0178 \ll 1$. Hence, it is the elastic interaction that dominates the motions of skyrmions instead of the backflow effect, which can be neglected.

The total free energy $F_{\text{total}}$ of the system is given by

$$F_{\text{total}} = \int_V (f_{\text{LdG}} + f_{el} + f_{\text{diel}} + f_{\text{flex}})dV + \int_S f_{\text{surf}} dS, \tag{1}$$

Where $f_{\text{LdG}}$ is the short-range Landau-de Gennes free energy, $f_{el}$ is the long-range elastic energy, $f_{\text{diel}}$ is the dielectric energy, $f_{\text{flex}}$ is the flexoelectricity energy, and $f_{\text{surf}}$ is the surface anchoring-induced free energy.

The skyrmions responds asymmetrically under opposite bias voltages. In the case of a dielectric response in LC under an electric field, the orientation of the director quadratically depends on the electric field. It means that the director does not differentiate the directionality of the electric field but instead chooses the fastest way to align either parallel or antiparallel to the field[26,27]. Hence, an additional polarization **P** response emerges in the system, exhibiting a linear coupling between the polarization and the electric field **E**. In traditional nematic materials, the spatial polarization arises from the flexoelectricity effects[28,29]. Regarding for the flexoelectric energy, the electric polarization is coupling to splay and bend deformations of the director fields, termed flexoelectricity:

$$\mathbf{P}_f = e_{11}(\mathbf{n} \cdot \nabla \cdot \mathbf{n}) - e_{33}[\mathbf{n} \times (\nabla \times \mathbf{n})], \tag{2}$$

where $e_{11}$ and $e_{33}$ are the usual flexoelectric coefficients, corresponding to splay and bend, seen in numerical simulation in supplementary information.

To validate the presence of the flexoelectric effect in our system, we use 5CB to test the mobility of soliton under the same experiment conditions, the skyrmions in which can only move at a low speed in the range of 10°. The response of the polarization to the AC field is relatively slow due to the significant reorientation of the polarization compared to the dielectric response[30,31]. Consequently, when applying a high-frequency AC field, the flexoelectric effect is significantly weakened and the asymmetrical motion of the solitons disappears.

Continuum simulations also reveal that the flexoelectricity effect is a key factor to induce the omnidirectional motions of skyrmions. In the simulations, an AC electric field with an amplitude of $E_0 = 0.48$ is applied. When the system exhibits a flexoelectric effect, under no bias electric field $E_b = 0$, the solitons move along the *y* direction, Fig. S3b. However, when excluding the flexoelectric effect, under no bias electric field, the soliton cannot move, Fig. S3a. This suggests that the velocity along the *y*-direction is induced by the oscillation of the flexo-polarization under the symmetry AC field. The electrical torque acting on the polarizations by the electric field can induce the reorientation of directors and generate non-symmetric director fields. Polarization reversal occurs when the direction of electric field reverses. The director fields are deformed periodically due to the polarization reversion, Fig. 3 a-h. In this case, when the bias electric field $E_b > 0$, the velocity along the positive *x*-direction increases and the motion direction $\alpha > 0$. While $E_b < 0$, the velocity along the negative *x*-direction increases and the motion direction $\alpha < 0$, supplementary Figs S4 and S5.

In order to better describe the motion of skyrmions, we captured the morphological changes of the skyrmions with a high-speed camera as a function of applied voltage. When the electric field frequency was 4 Hz and the camera frame rate was 1000 fps, the morphological changes of the skyrmions within one voltage cycle could be observed, Fig. 3i. The intensity of the director far away from the skyrmion alternated between bright and dark over time, which means transitions from the vertical to tilted and then in-plane orientation. The director field of the skyrmions also responded to the fluctuation to minimize the total free energy, resulting in morphological changes within one voltage cycle. Numerically comparing the director fields of the skyrmion within one period of the AC field, when the system excludes the flexoelectric effect, due to the dielectric response, the skyrmion can maintain high degree of rotational symmetry and the helicity remains around $\pi/2$, Fig. S6 a. In the presence of the flexoelectric effect, the polarization response results in a symmetry breaking of the skyrmion shape. The skyrmion elongates along rubbing directions and its helicity deviates from $\pi/2$, Fig. 3j.

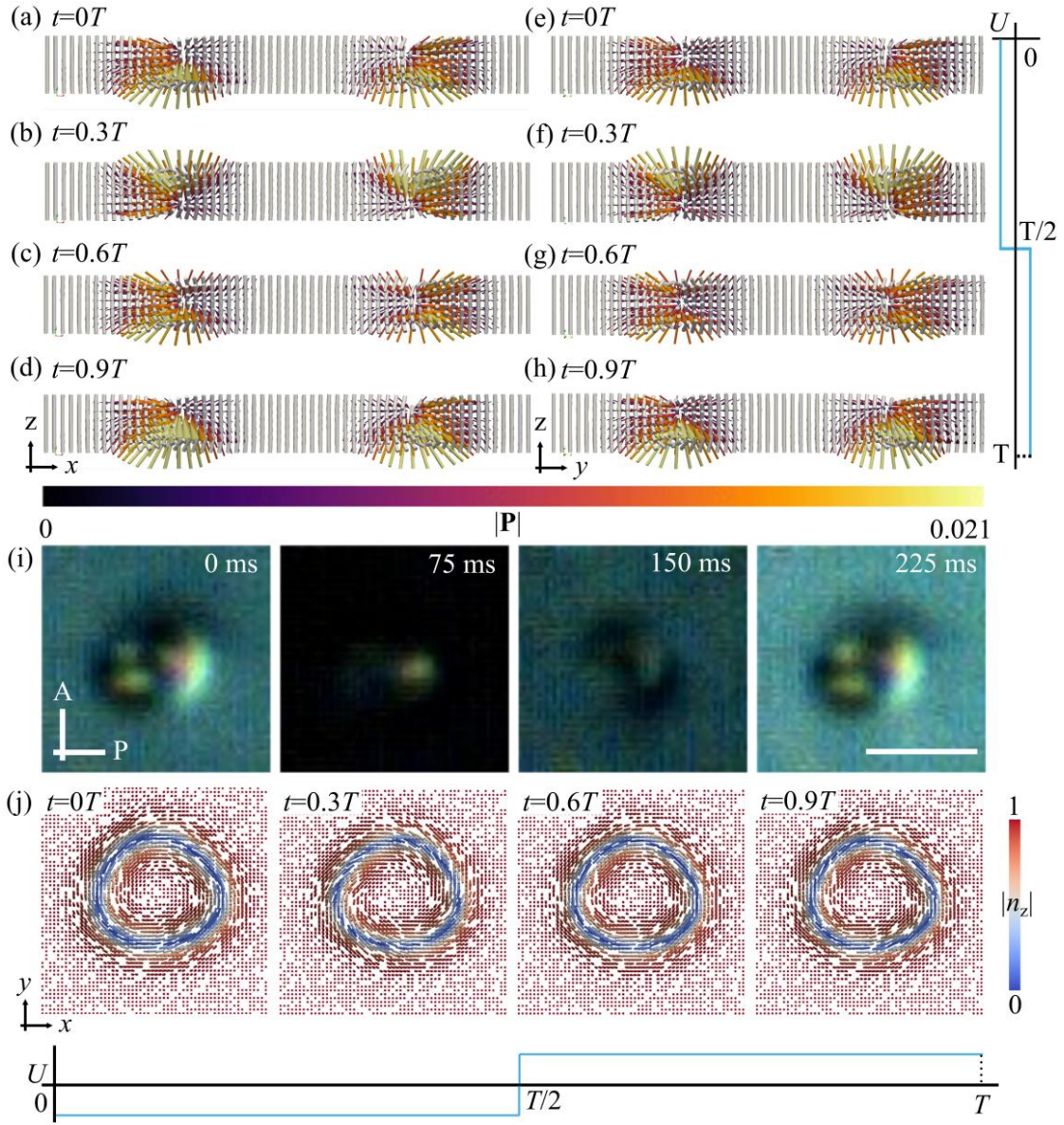

**Fig 3. Variation of skyrmions in one period of AC fields. a-h,** Variation of director fields of skyrmions and polarization fields in *zx*- and *zy*-cross-sectional in one period of AC fields. **i,** POM micrographs of a skyrmion within one voltage cycle. **j,** Variation of *xy*-cross-sectional director fields of skyrmions in one period of AC fields. Color indicates the magnitude of the director field $|n_z|$ in the *z*-direction.

**Structure change of skyrmions influenced by electric field.**

As $U_b$ increases, the solitons transform from a compact quadrifoliate shape to a long loop shape, Fig. 4a. We observed and measured the morphology of the skyrmion when the voltage reached its maximum value. The size of the skyrmions along the *x*-axis is referred to as $L_x$, while the width along the y-axis is denoted as $L_y$. Both $L_x$ and $L_y$ remain almost unchanged when $|U_b| < 5\,\text{V}$, Fig. 4b. The size of the solitons increases when $|U_b| > 5\,\text{V}$, and the ratio of $L_x / L_y$ is approximately $1.5 \sim 2.0$.

Actually, the size of the skyrmion changes periodically with the voltage cycle. The maximum deformation along the x-direction shows a difference of approximately $3\,\mu\text{m}$ between the maximum and minimum sizes, while the size variation along the y-direction is approximately $1\,\mu\text{m}$, Fig 4c.

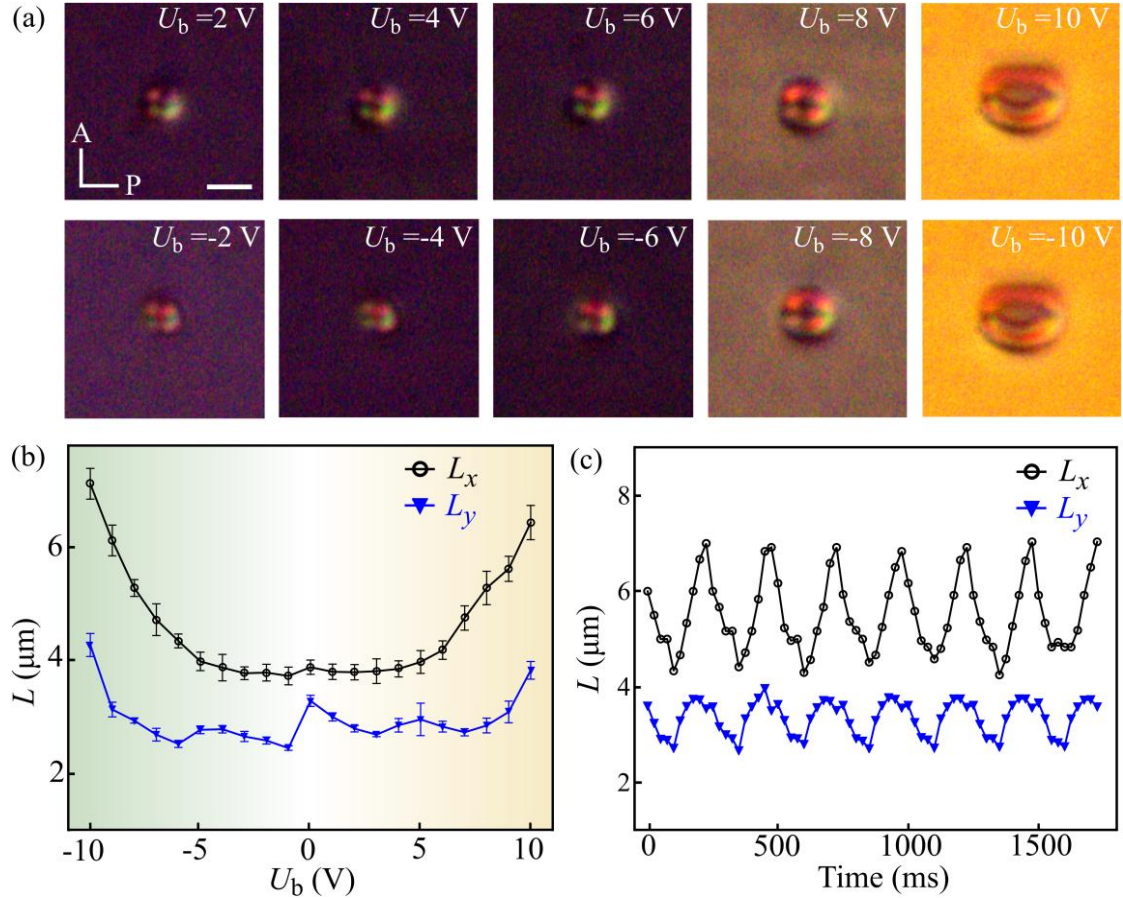

**Fig. 4 The changes of the skyrmions influenced by electric field. a,** Polarization microscopy images of a skyrmion at different bias voltages. Scale bar $5\,\mu\text{m}$. **b,** Dependence of mean size, $L_x$ and $L_y$, of topological solitons on the applied bias voltage, $U_b$. **c,** Dependence of mean size, $L_x$ and $L_y$, of a skyrmion within one voltage cycle.

**Electric field control of trajectories**

Another striking feature is that skyrmions' motion directions can promptly respond to the changes in the electric field. Thus, abruptly adjusting the bias voltage can change the orientation to any desired angle, as shown in Fig. 5a and Supplementary Movie 6. By changing $U_b$ from 0 to 2.0 V, 7.0 V, and 11.0 V, at $f = 20$ Hz we can achieve direction changes of $-\pi/4$, $-\pi/2$, $-3\pi/4$, and $-\pi$, respectively.

By harnessing the controllability of skyrmions' motion directions, it becomes feasible to create more complex patterns through the switching of the bias voltage. By capturing the position of individual solitons at regular intervals and plotting these positions on an image, the motion trajectories of the words "NJUPT" and "HKUST" are presented in Figs. 5b, c, and Supplementary Movie 7.

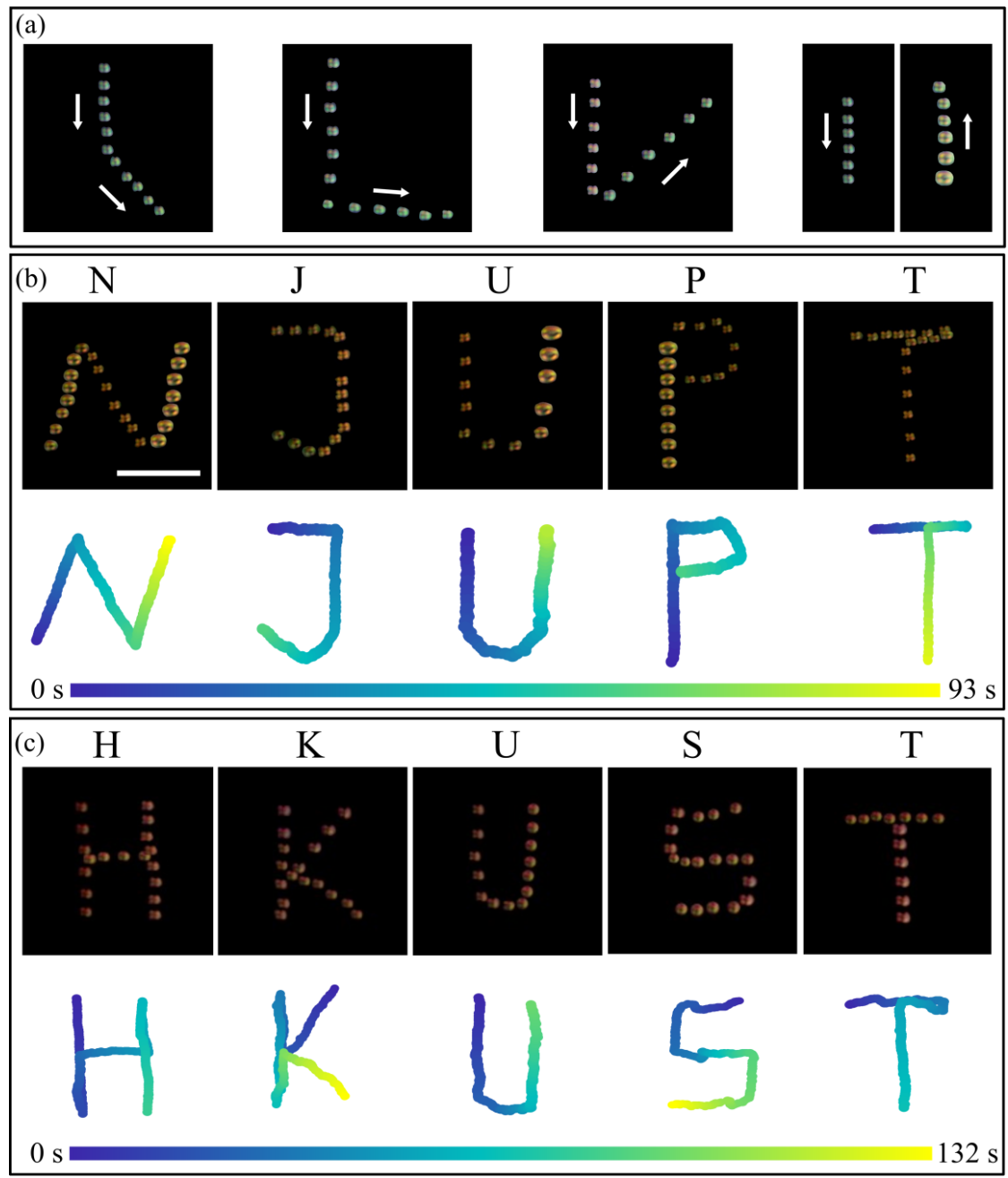

**Fig. 5 Manipulation of trajectory of skyrmions. a,** Direction changes of $-\pi/4$, $-\pi/2$, $-3\pi/4$, and $-\pi$ of skyrmions by switching the voltage. **b-c,** Motion along the "NJUPT" and "HKUST" trajectories, with corresponding color gradient trajectory maps. Scale bar: 10 μm.

**Conclusions**

To summarize, we demonstrated 360°-tunable motion of topological soliton "skyrmions" in a chiral nematic liquid crystal powered by an AC square wave with a bias voltage. The size, velocity, and population density of the solitons depend strongly on the bias voltage. By adjusting the magnitude of the bias voltage, it is possible to drive the skyrmions to move along any predetermined trajectory. The 360°-tunable motion of the skyrmions do not require a high temperature. Numerical modeling gives insights into this field-driven omnidirectional skyrmions. A spatial polarization arising from the flexoelectricity effects emerges in the system, exhibiting a linear coupling with the electric field. The flexoelectric-polarization leads to an asymmetrical response of skyrmions under opposite bias voltages, which is also a key factor to induce the omnidirectional motions of skyrmions. Polarization reversal occurs when the direction of electric field reverses, resulting in a periodic deformation of the director fields. The oscillation of the flexo-polarization induces the velocity along the y-direction. When the bias voltage is positive, the velocity along the positive x-direction increases and the motion direction $\alpha > 0$. Conversely, if the bias voltage is negative, the velocity in the negative x-direction increases and the motion direction $\alpha < 0$. Therefore, the direction of the overall velocity can be altered by adjusting the bias voltage.

Different from the squirming motion of the skyrmions driven by modulated wave[15], this experiment breaks the structural symmetry of the skyrmions through planar anchoring, enabling unidirectional motion under the drive of a normal square wave. The applied bias voltage generates spatial polarization, allowing the skyrmions to move in any direction within a two-dimensional plane, a capability that non-topological solitons cannot achieve[5,21,22]. This is the first time that highly tunable motion of topological solitons has been observed, which also provides a theoretical basis for the study of topological defects in liquid crystals. Moreover, the topological solitons with tunable motion directions can be applied in various fields such as transportation of micro-cargos, beam-steering, and racetrack memories.

## Materials and methods

**Sample preparation.** The chiral nematic liquid crystals utilized in this study, which possess a pitch of approximately $p \sim 2.7\,\mu\text{m}$, measured using a stripe–wedge

Grandjean–Cano cell, Supplementary Fig. S7, are prepared by combining a nematic liquid crystal, DP002-026 (purchased from Jiangsu Hecheng Display Technology,) with a chiral dopant, S811 (purchased from Nanjing Leyao). DP002-026 exhibits a dielectric anisotropy of $\Delta\varepsilon=+4.8$ at $f=1\,\text{kHz}$. The nematic-isotropic phase transition temperature of DP002-026 is 108°C. To prepare the mixture, DP002-026 and S811 are individually dissolved in dichloromethane (DCM) and subsequently mixed together. The mass fractions of DP002-026 and S811 are 97 wt% and 3 wt%, respectively. The resulting solution is subjected to 20 minutes of thorough ultrasonic treatment in an ultrasonic cleaner at a temperature of 34°C. Subsequently, the mixture is heated and dried on a hot plate set at 80°C for 1 hour to allow for the evaporation of the DCM solvent. After the evaporation process, the prepared mixtures are injected into the cells at 120°C in their isotropic phase.

In order to observe the motions of topological solitons, the cells are composed of two glass plates with transparent indium tin oxide (ITO) electrodes of low resistivity (ranging between 10 and 50 Ω/sq). The inner surfaces of the glass plates are coated with polyimide PI-2555 (purchased from Nanjing Ningcui) at 2,700 rpm for 30 s, followed by a 5-min prebake at 120°C and a 20-min bake at 220°C, and then are unidirectionally rubbed to achieve a planar alignment of the chiral nematic liquid crystals, resulting in a helical pitch that is perpendicular to the glass substrates. To control the cell thickness within a range of $d=7.0\pm0.5\,\mu\text{m}$, silica spheres are dispersed in UV glue. Two wires are led out from the upper and lower substrates of the liquid crystal cell to provide electric field.

**Generation of solitons.** The electric field is applied perpendicularly to the chiral nematic liquid crystal cells at 34°C on a hot stage (HCS402XY, Instec) using a waveform generator (AFG1022, Agilent) and a high voltage amplifier (ATA-2041, Aigtek). The voltage values are measured using a multimeter (DMM6500, Keithley).

**Optical characterization.** Samples and the motions of topological solitons are observed using a polarizing microscope (ECLIPSE Ci-POL). Images and videos are recorded using a digital camera (E3ISPM09000KPB, Kainuo, China) and a high-speed camera (SH3-502, SinceVision, China). During the motion measurements, the frame rate of the digital camera is adjusted to 20 frames per second to ensure capturing the same moment within each voltage cycle.

**Data analysis.** The motion of the solitons is tracked and analyzed by using open-source ImageJ and Fiji software. The position information and number density of the solitons are extracted for each frame through a plugin of ImageJ[32], TrackMate. The temporal evolution of velocity order is obtained by analyzing the positional information of solitons between frames of movies. Then, the data analysis and plotting in MATLAB (obtained from MathWorks) and Mathematicas software are then performed to characterize velocity, motion direction, and density.

**Numerical Simulation.** The total free energy $F_{total}$ of a nematic liquid crystal is given by

$$F_{total} = \int_V (f_{LdG} + f_{el} + f_{diel} + f_{flex})dV + \int_S f_{surf}dS \quad (1)$$

where $f_{LdG}$ is the short-range Landau-de Gennes free energy, $f_{el}$ is the long-range elastic energy, $f_{diel}$ is the dielectric energy, $f_{flex}$ is the flexoelectricity energy, and $f_{surf}$ is the surface anchoring-induced free energy. The Landau-de Gennes free energy density $f_{LdG}$ is calculated as

$$f_{LdG} = \frac{A_0}{2}(1-\frac{U}{3})\text{Tr}(\mathbf{Q}^2) - \frac{A_0 U}{3}\text{Tr}(\mathbf{Q}^3) + \frac{A_0 U}{4}[\text{Tr}(\mathbf{Q}^2)]^2 \quad (2)$$

where $\mathbf{Q}$ is the tensorial order parameter from an ensemble average over unit vector $\mathbf{n}$ (representing the molecular orientation), $\mathbf{Q} = \langle \mathbf{nn} - \mathbf{I}/3 \rangle$. Parameter $U$ controls the magnitude of $S$ of a homogenous static system through

$$S = \frac{1}{4} + \frac{3}{4}\sqrt{1-\frac{8}{3U}} \quad (3)$$

The Frank-Oseen elastic free energy density for the nematic liquid crystal is expressed as[33]

$$\begin{aligned}f_{el}^{FO} =& \frac{1}{2}K_1(\nabla \cdot \mathbf{n})^2 + \frac{1}{2}K_2(\mathbf{n}\cdot\nabla\times\mathbf{n}+q_0)^2 \\ &+ \frac{1}{2}K_3[\mathbf{n}\times(\nabla\times\mathbf{n})]^2 - \frac{1}{2}K_{24}\nabla\cdot[\mathbf{n}(\nabla\cdot\mathbf{n})+\mathbf{n}\times(\nabla\times\mathbf{n})]\end{aligned} \quad (4)$$

where $K_1$, $K_2$, $K_3$ and $K_{24}$ are the splay, twist, bend and saddle-slay elastic moduli, respectively. To be consistent with the Landau-de Gennes energy, the elastic energy density in the simulation is alternatively written in terms of the **Q** tensor (in the Einstein summation form)

$$f_{el}^Q = \frac{1}{2}L_1(\partial_k Q_{ij})(\partial_k Q_{ij}) + \frac{1}{2}L_2(\partial_k Q_{jk})(\partial_l Q_{jl}) + \frac{1}{2}L_3 Q_{ij}(\partial_i Q_{kl})(\partial_j Q_{kl}) \\ + \frac{1}{2}L_4(\partial_l Q_{jk})(\partial_k Q_{jl}) + L_5 q_0 \varepsilon_{ikl} Q_{ij} \partial_k Q_{lj} \tag{5}$$

The mapping between constant sets $K_1$, $K_2$, $K_3$, and $K_{24}$ and $L_1$, $L_2$, $L_3$, $L_4$ and $L_5$ is via

$$L_1 = \frac{1}{2S^2}[K_{22} + \frac{1}{d}(K_{33} - K_{11})]$$
$$L_2 = \frac{1}{S^2}(K_{11} - K_{22} - K_{24})$$
$$L_3 = \frac{1}{2S^3}(K_{33} - K_{11}) \tag{6}$$
$$L_4 = \frac{1}{S^2}K_{24}$$
$$L_5 = \frac{1}{S^2}K_{22}$$

The dielectric energy contribution is given by

$$f_{diel} = -\frac{1}{2}\varepsilon_0 \varepsilon_{ij} E_i E_j \tag{7}$$

where $\varepsilon_0$ represents the dielectric permittivity of vacuum, and $\varepsilon_{ij}$ corresponds to the tensorial dielectric permittivity of the nematic material:

$$\varepsilon_{ij} = \varepsilon_\perp \delta_{ij} + (\varepsilon_\parallel - \varepsilon_\perp)n_i n_j \tag{8}$$

Here, $\varepsilon_\perp$ and $\varepsilon_\parallel$ are the dielectric permittivities perpendicular and parallel to the nematic field, respectively. By introducing an isotropic dielectric permittivity $\bar{\varepsilon} = (2\varepsilon_\perp + \varepsilon_\parallel)/3$ and a permittivity anisotropy $\varepsilon_a = 2(\varepsilon_\parallel - \varepsilon_\perp)/3$, one has $\varepsilon_{ij} = \bar{\varepsilon}\delta_{ij} + \varepsilon_a Q_{ij}$. The above expressions can be rewritten in terms of the Q-tensor as

$$f_{diel} = -\frac{1}{2}\varepsilon_0(\bar{\varepsilon}\delta_{ij} + \varepsilon_a Q_{ij})E_i E_j \tag{9}$$

Regarding for the flexoelectric energy, the electric polarization is coupling to splay

and bend deformations of the director fields, termed flexoelectricity:

$$\mathbf{P}_f = e_{11}(\mathbf{n} \cdot \nabla \cdot \mathbf{n}) - e_{33}[\mathbf{n} \times (\nabla \times \mathbf{n})] \tag{10}$$

where $e_{11}$ and $e_{33}$ are the usual flexoelectric terms, corresponding to splay and bend.

We can also express the flexoelectric polarizations in tensor form[34,35]:

$$P_i = R_1 \partial_\gamma Q_{i\gamma} + R_5 Q_{\beta\gamma} \partial_\gamma Q_{i\beta} + R_6 Q_{i\alpha} \partial_\gamma Q_{\alpha\gamma} \tag{11}$$

The flexoelectric term can be described in tensor form:

$$f_{flex} = -E_i P_i = -E_i (R_1 \partial_\gamma Q_{i\gamma} + R_5 Q_{\beta\gamma} \partial_\gamma Q_{i\beta} + R_6 Q_{i\alpha} \partial_\gamma Q_{\alpha\gamma}) \tag{12}$$

The coefficient can be converted to:

$$\begin{aligned} e_{11} &= \frac{3}{2} R_1 S + \frac{3}{4} (2R_6 - R_5) S^2 \\ e_{33} &= \frac{3}{2} R_1 S + \frac{3}{4} (2R_5 - R_6) S^2 \end{aligned} \tag{13}$$

In our case, we are dealing with the non-singular topological structures, so we use the parameter of equilibrium $S_{eq} = 0.63$. We can set the flexoelectricity coefficient in our simulations:

$$e_{33} - e_{11} = \frac{9}{4} S_{eq}^2 (R_5 - R_6) \tag{14}$$

In our system, the asymmetry response is induced by the flexoelectric polarization. However, to induce the motion of soliton, the dielectric effect dominates the system. Thus, we adopt a simplified model that assumes a uniform electric field **E** and does not integrate Maxwell's equations with the liquid crystal structure in our simulations.

The anchoring energy $f_{surf}$ is calculated by the nondegenerate formula, the so-called the Rapini-Papoular[36]

$$f_{surf} = \frac{1}{2} W (\mathbf{Q} - \mathbf{Q}_{surf})^2 \tag{15}$$

where $\mathbf{Q}_{surf}$ is the preferred field of the surface, $\mathbf{Q}_{surf} = S(\mathbf{n}_s \mathbf{n}_s - \mathbf{I}/3)$, $\mathbf{n}_s$ is the surface-preferred molecular orientation and $W$ is the anchoring strength. In our simulations, we assumed that the anchoring conditions on the top and bottom surfaces are strong, and therefore, the directors are almost fixed at the substrates.

Eq. (5) calculates a thermodynamic potential that determines the stable or metastable solutions of the system. We define a molecular field

$$\mathbf{H} = -[\frac{\delta F_{total}}{\delta \mathbf{Q}}]^{st} \tag{16}$$

where $[\frac{\delta F_{total}}{\delta \mathbf{Q}}]^{st}$ is a symmetric and traceless operator. Assuming that all transitions are quasistatic processes, the evolution of the **Q**-tensor for bulk points follows

$$\partial_t \mathbf{Q} = \Gamma_S \mathbf{H} \quad (17)$$

where $\Gamma_S$ is the relaxation constant. For surface points, the evolution of **Q** is governed by

$$\frac{\partial \mathbf{Q}}{\partial t} = -\Gamma_S [-v \cdot \frac{\partial F}{\partial \nabla \mathbf{Q}} + \{\frac{\partial f_{surf}}{\partial \mathbf{Q}}\}^{st}] \quad (18)$$

where the unit vector $v$ represents the surface normal.

The numerical parameters used here are $A_0 = 1.17 \times 10^5$ J/m$^3$ and $U = 3.5$. We use one elastic constant in our simulations, where $K = 10^{-11}$ N. The dielectric anisotropic of the material used here is 1 and the flexoelectric coefficients are $R_5 - R_6 = 0.5$. In terms of the AC field, the period is set to be 500 simulation steps, and the amplitude is $U_0 = 0.48$.

The size of the simulation box is $[L_x, L_y, L_z] = [300, 300, 120]$. The intrinsic pitch is $p_0 = 50$ and the chirality of the liquid crystal is $q_0 = 2\pi/50 = 0.13$. The boundary conditions in x- and y- directions are periodic boundary conditions while the boundary condition in the z-direction is finite anchoring conditions with anchoring strength W=$1 \times 10^{-4}$ J/m$^2$.

## Acknowledgments

The work is supported by the National Key Research and Development Program of China (No.2022YFA1405000), the National Natural Science Foundation of China (No. 62375141), the Guangdong Natural Science Foundation (No. 2022A1515011186), the Natural Science Foundation of Jiangsu Province, Major Project (No. BK20243067).

## Author contributions

J.H.C. and W.T.T. contributed equally to this work. J.H.C., J.N., and Y.D. performed the experiments. W.T.T. performed the theoretical analysis and numerical simulations. J.H.C., J.N., and Y.D. analyzed data. B.X.L., R.Z., J.J.d.P. and Y.Q.L. conceived, designed, and directed the research. J.H.C., W.T.T., Y.D., Y.X.C, B.X.L., R.Z., J.J.d.P., and Y.Q.L. contributed to the discussion and wrote the manuscript.

## Data availability

The data that support the findings of the study are available from the correspondingauthor upon reasonable request.

## Competing interests

The authors declare no competing interests.

**Supplementary Figure S1**

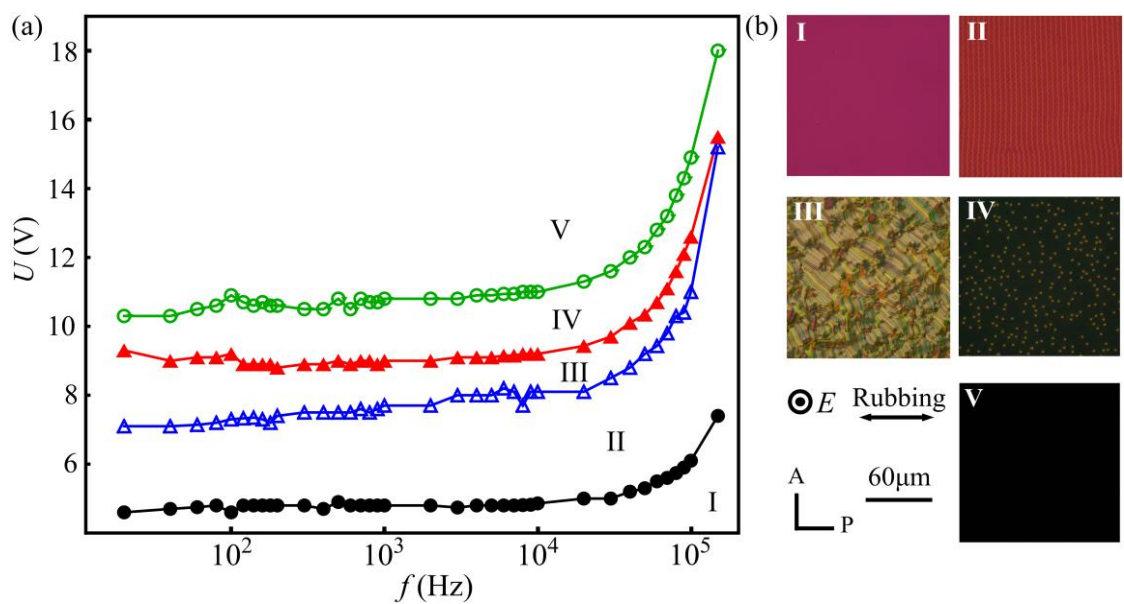

**Fig. S1 a,** frequency dependence of the threshold voltages for **b,** different states (I, helical state; II, rolls pattern; III, fingerprint textures; IV, solitons; V, Homotropic state) at $T = 34°C$ and corresponding polarizing micrographs of different CNLC states are observed between crossed polarizers at various voltages of frequency $f = 20$ Hz. The cell thickness is $d = 6.7$ μm. Scale bar 60 μm.

**Supplementary Figure S2**

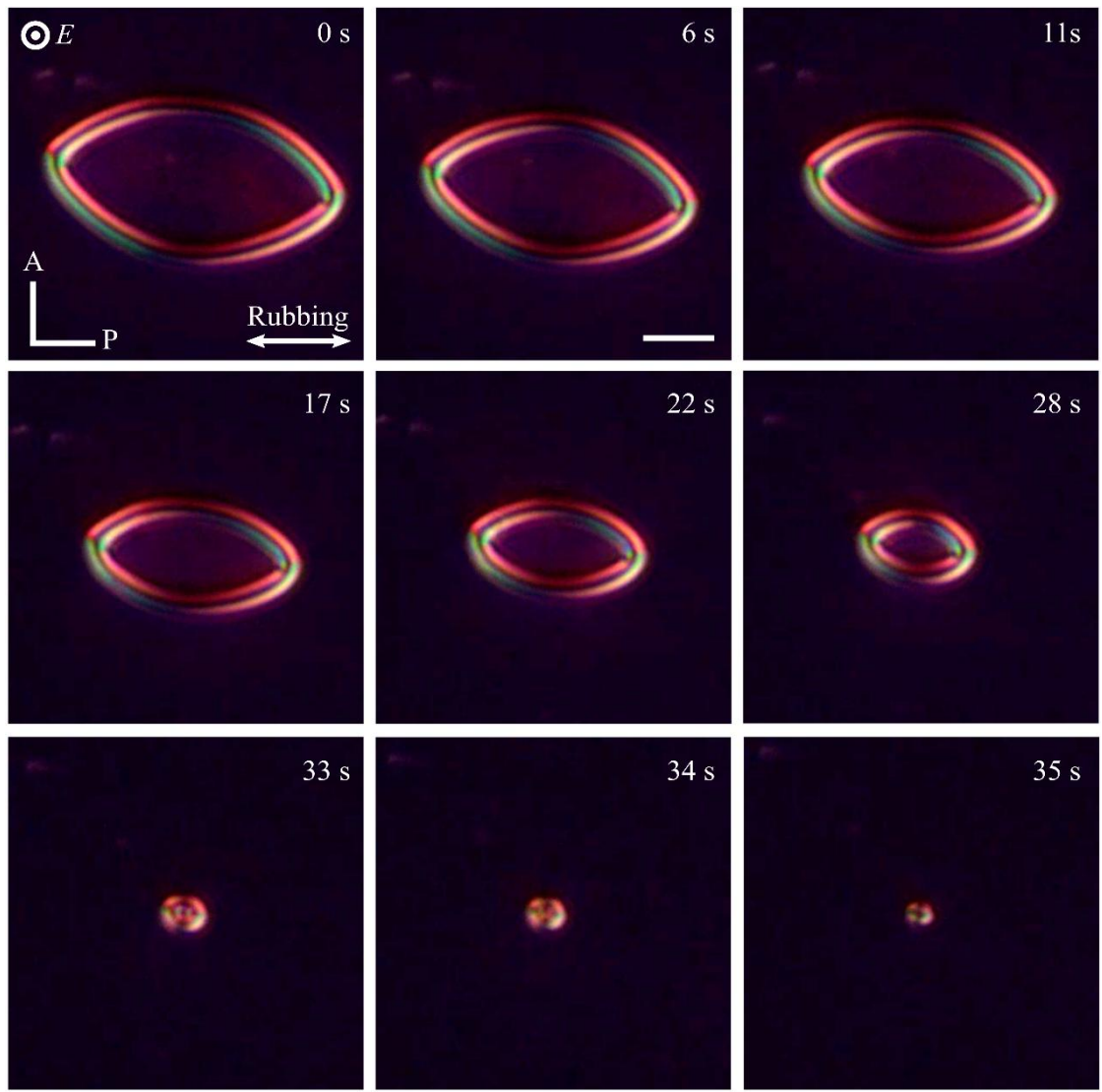

**Fig. S2** Collapse of a looped cholesteric finger into a soliton. The cell thickness is 6.5 μm. $T = 34°C$. $f = 20$ Hz, $U_b = 0$. Scale bar 15 μm.

**Supplementary Figure S3**

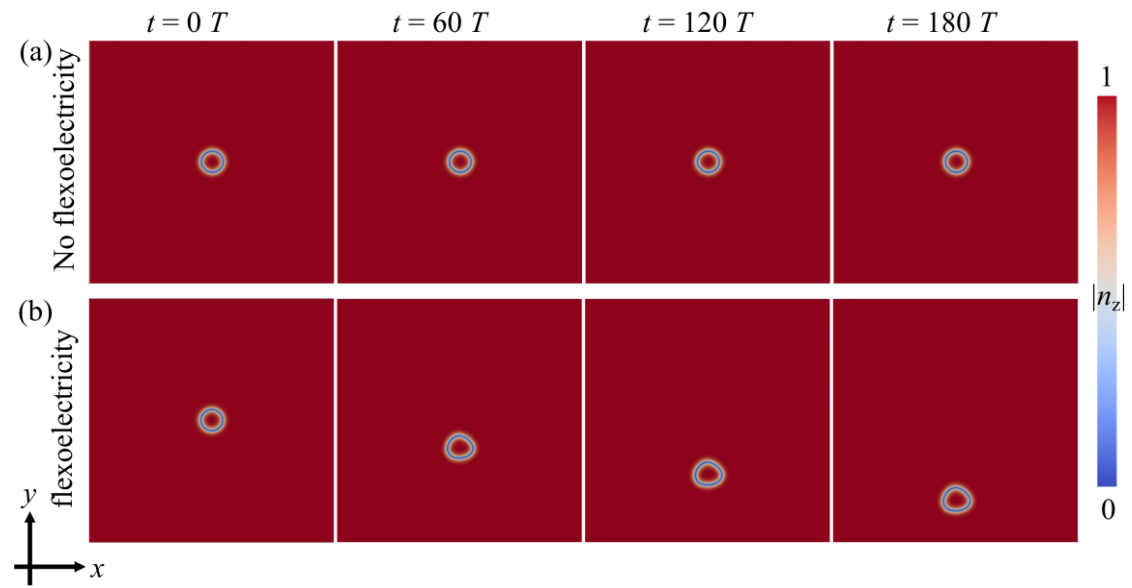

**Fig. S3** The positions of skyrmion over time under AC field with no bias voltages. The flexoelectricity effects are excluded in **a** and included in **b**. Color indicates the magnitude of the director field $|n_z|$ in the $z$- direction.

**Supplementary Figure S4**

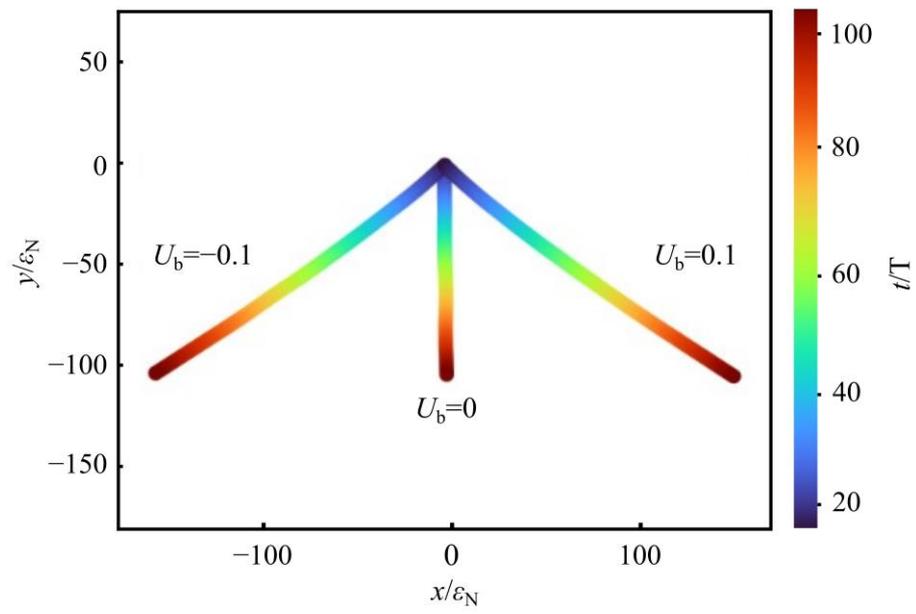

**Fig. S4** The movement trajectories of solitons simulated under different bias voltages.

**Supplementary Figure S5**

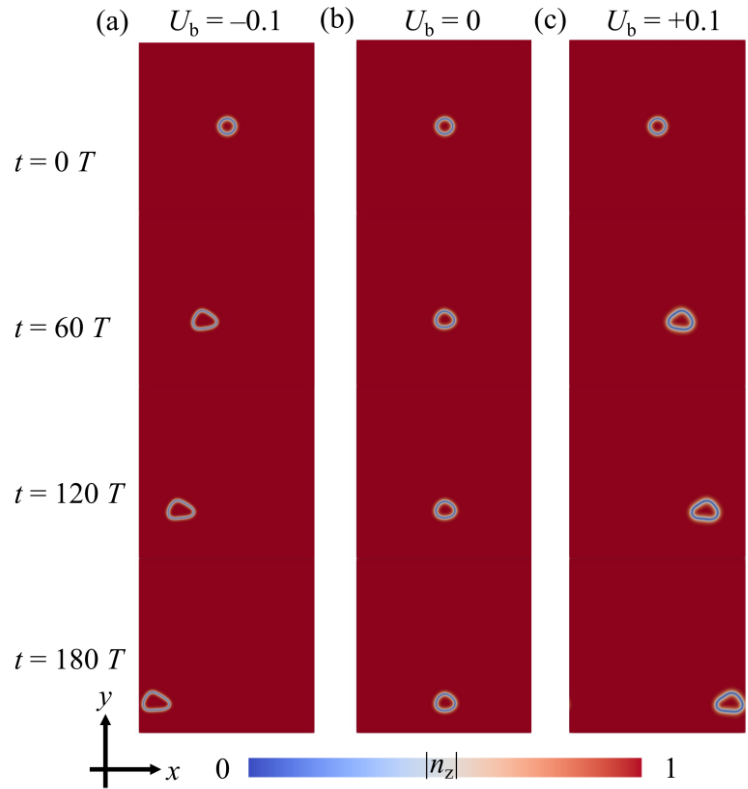

**Fig. S5** The skyrmion positions over time under AC field with different bias voltages. Color indicates the magnitude of the director field $|n_z|$ in the z- direction. **a,** The skyrmion under bias voltages of $U_b = -0.1$ translates along the direction of $-40°$. **b,** The skyrmion under no bias voltages of $U_b = 0$ translates along the direction of $-90°$. **c,** The skyrmion under bias voltages of $U_b = 0.1$ translates along the direction of $40°$.

**Supplementary Figure S6**

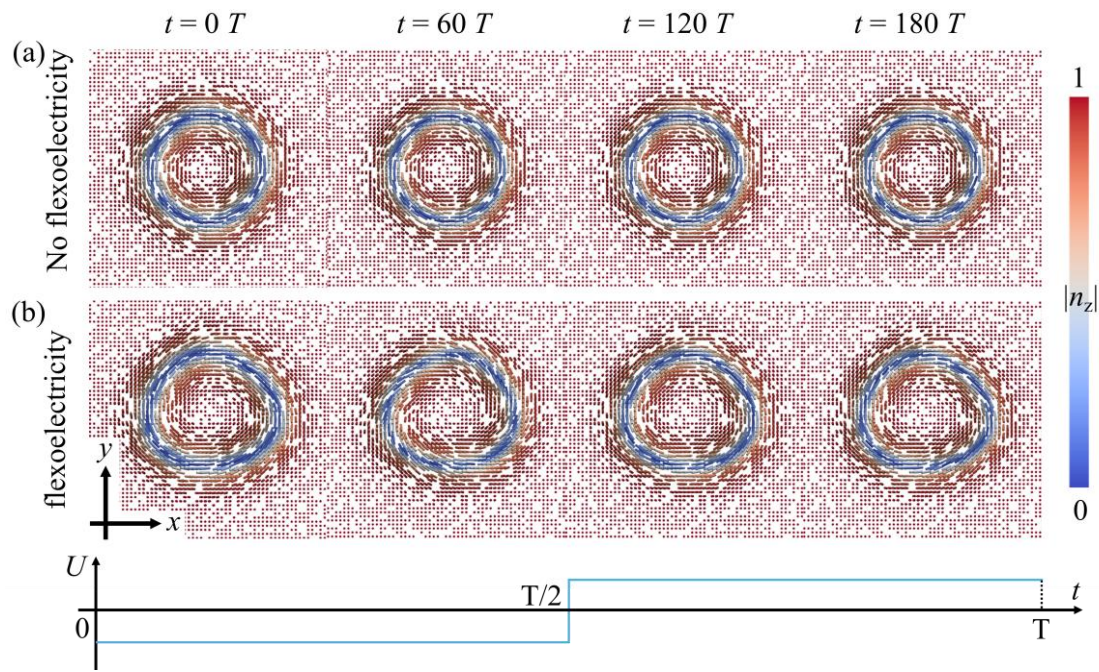

**Fig. S6** Variation of *xy*-cross-sectional director fields of skyrmions in one period of AC fields. Color indicates the magnitude of the director field $|n_z|$ in the *z*- direction. The flexoelectricity effects are excluded in **a** and included in **b**.

**Supplementary Figure S7**

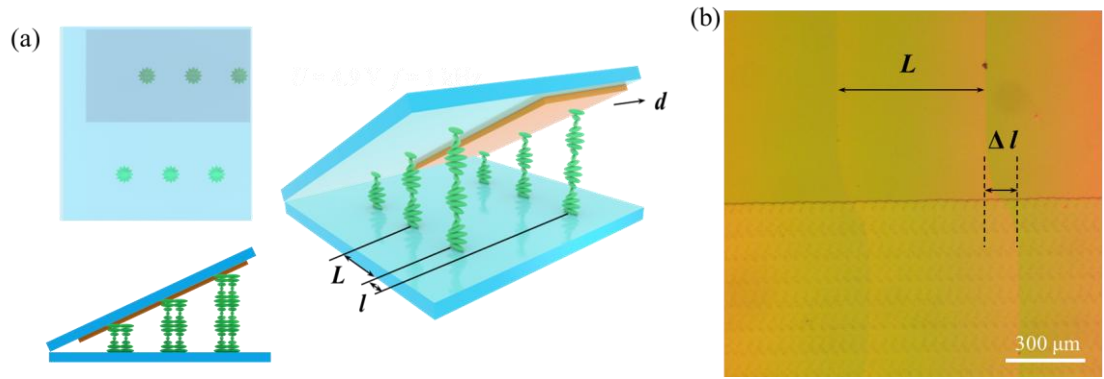

**Fig. S7 a,** Illustration of the stripe–wedge Grandjean–Cano cell. **b,** Microphotograph of a stripe–wedge Grandjean–Cano cell filled with chiral nematic liquid crystal. The cell thickness gradually increases from left to right. The upper half represents the area without ITO, while the lower half represents the area with ITO.

*Supplementary Information*

# Omnidirectionally manipulated skyrmions in an orientationally chiral system

Jiahao Chen, Wentao Tang, Xingzhou Tang, Yang Ding, Jie Ni, Yuxi Chen, Bingxiang Li, Rui Zhang, Juan de Pablo, Yanqing Lu

## Table of contents

**Supplementary Figure S1**

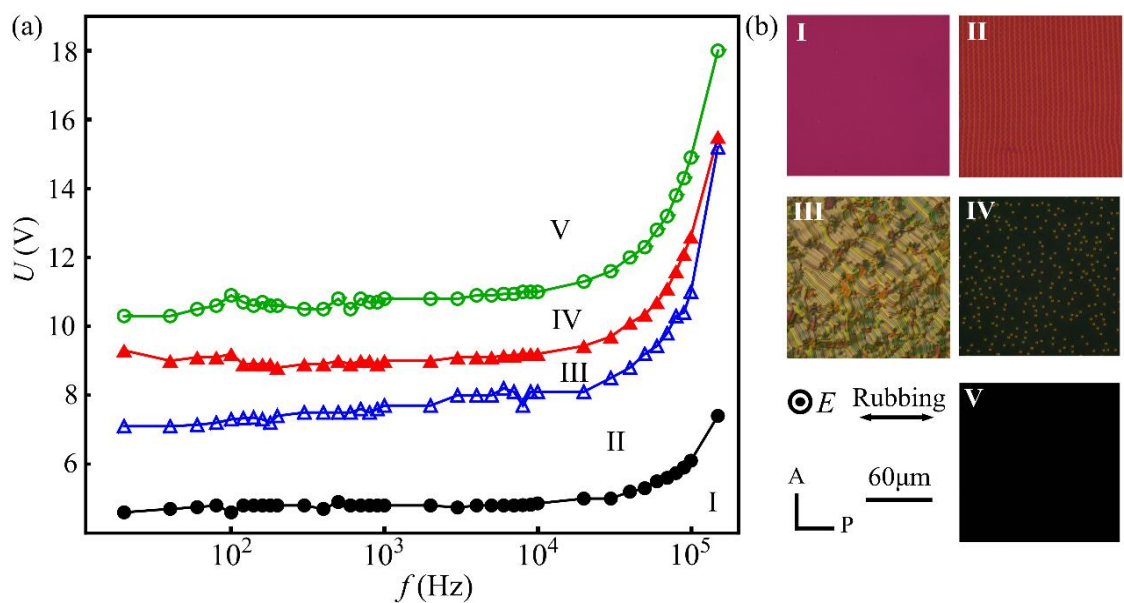

**Fig. S1 a,** frequency dependence of the threshold voltages for **b,** different states (I, helical state; II, rolls pattern; III, fingerprint textures; IV, solitons; V, Homotropic state) at $T = 34°C$ and corresponding polarizing micrographs of different CNLC states are observed between crossed polarizers at various voltages of frequency $f = 20\,\text{Hz}$. The cell thickness is $d = 6.7\,\mu\text{m}$. Scale bar $60\,\mu\text{m}$.

**Supplementary Figure S2**

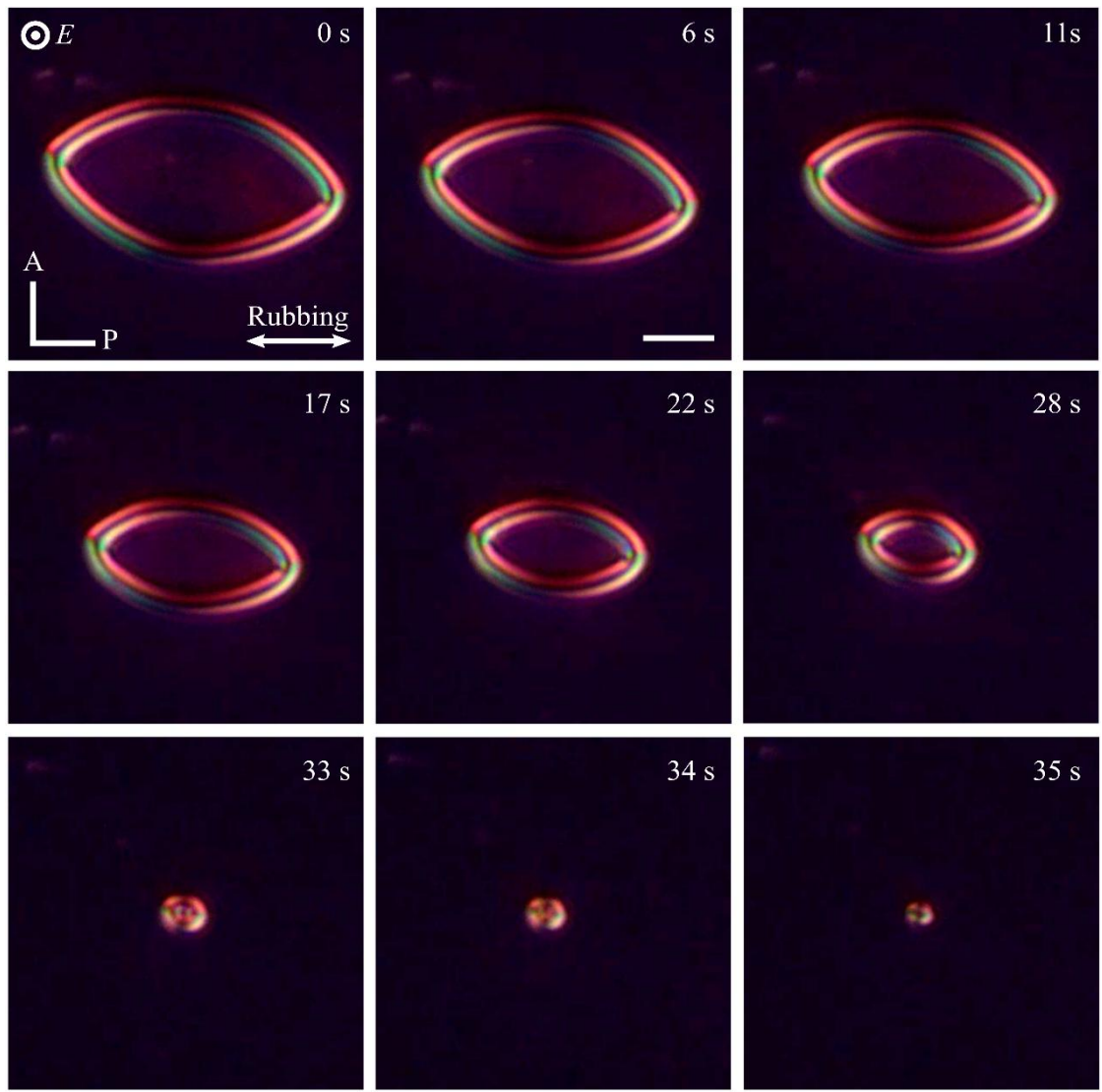

**Fig. S2** Collapse of a looped cholesteric finger into a soliton. The cell thickness is 6.5 μm. $T = 34°C$. $f = 20$ Hz, $U_b = 0$. Scale bar 15 μm.

**Supplementary Figure S3**

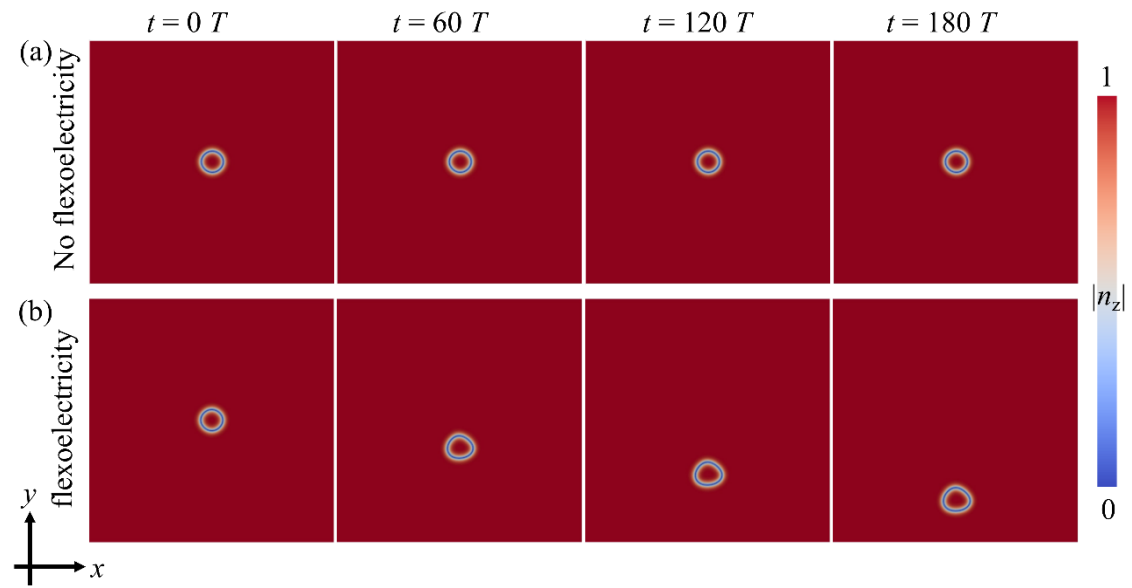

**Fig. S3** The positions of skyrmion over time under AC field with no bias voltages. The flexoelectricity effects are excluded in **a** and included in **b**. Color indicates the magnitude of the director field $|n_z|$ in the $z$- direction.

**Supplementary Figure S4**

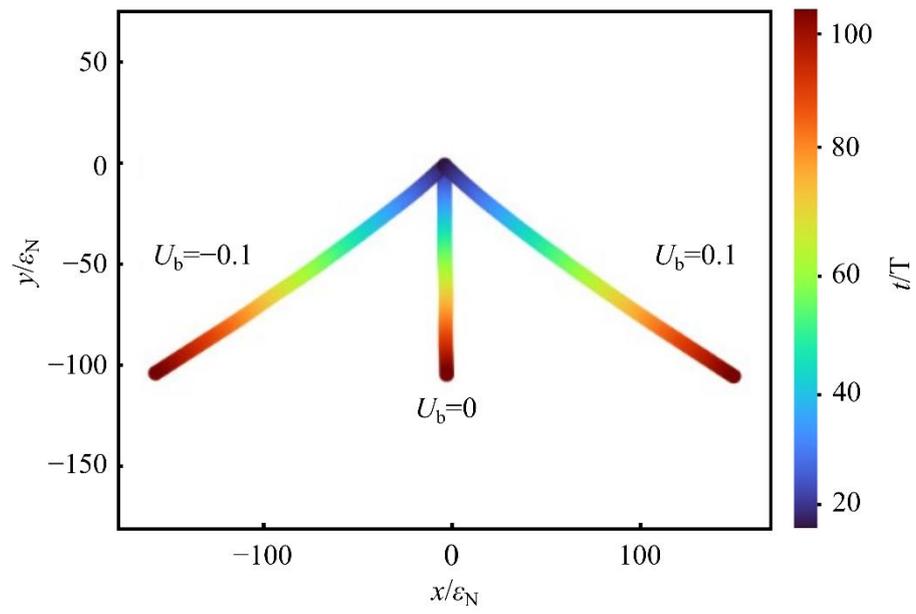

**Fig. S4** The movement trajectories of solitons simulated under different bias voltages.

**Supplementary Figure S5**

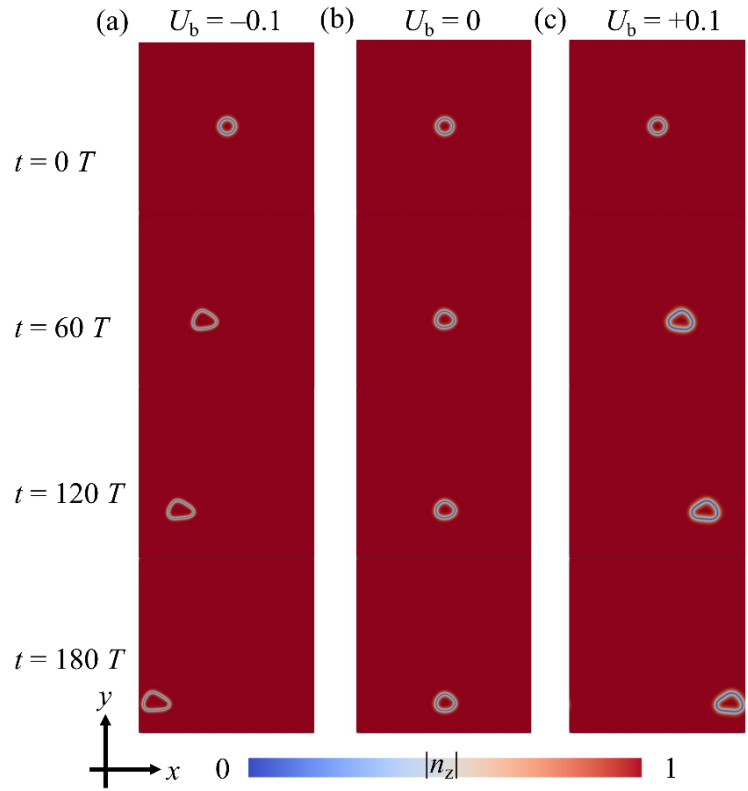

**Fig. S5** The skyrmion positions over time under AC field with different bias voltages. Color indicates the magnitude of the director field $|n_z|$ in the $z$- direction. **a,** The skyrmion under bias voltages of $U_b = -0.1$ translates along the direction of $-40°$. **b,** The skyrmion under no bias voltages of $U_b = 0$ translates along the direction of $-90°$. **c,** The skyrmion under bias voltages of $U_b = 0.1$ translates along the direction of $40°$.

**Supplementary Figure S6**

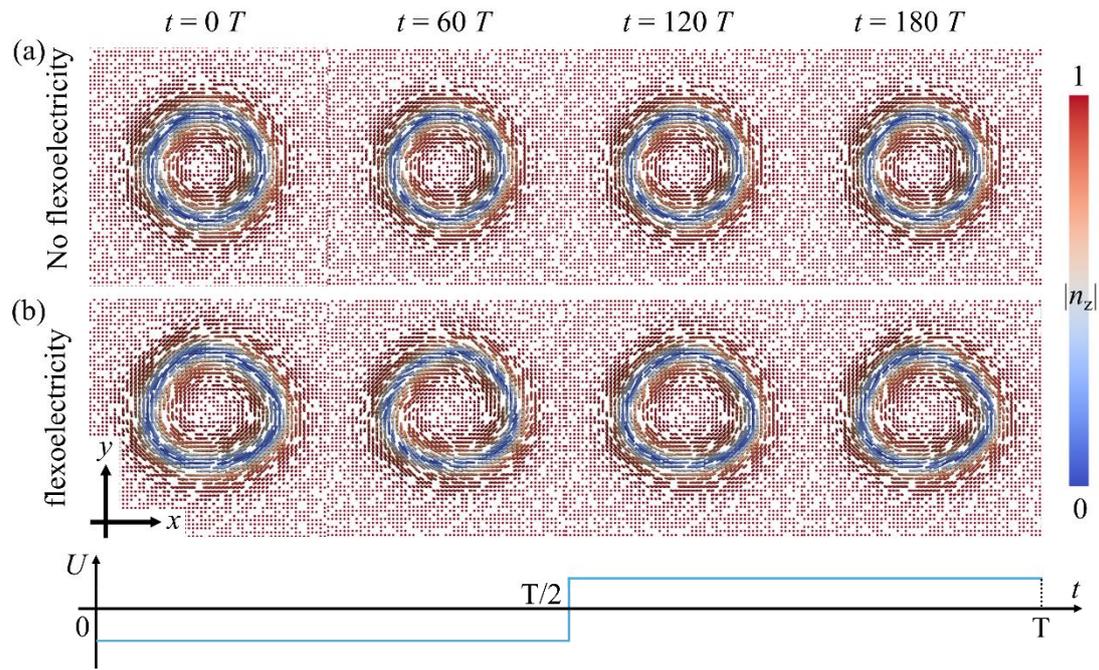

**Fig. S6** Variation of *xy*-cross-sectional director fields of skyrmions in one period of AC fields. Color indicates the magnitude of the director field $|n_z|$ in the *z*- direction. The flexoelectricity effects are excluded in **a** and included in **b**.

**Supplementary Figure S7**

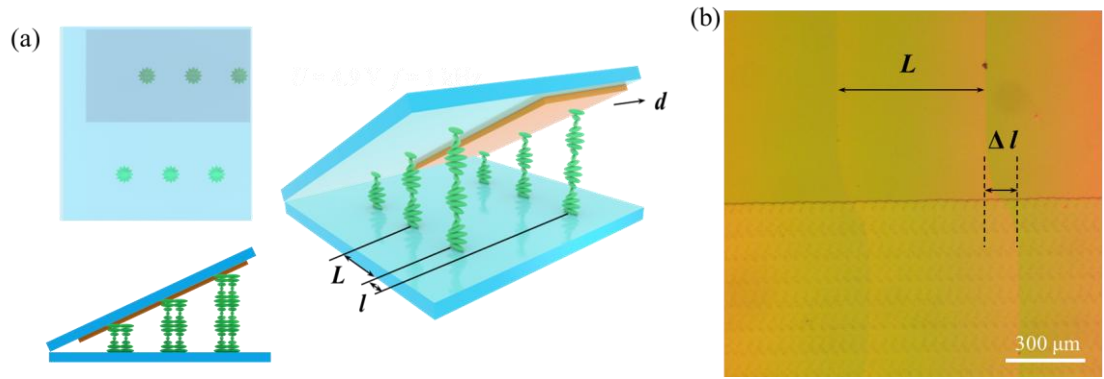

**Fig. S7 a,** Illustration of the stripe–wedge Grandjean–Cano cell. **b,** Microphotograph of a stripe–wedge Grandjean–Cano cell filled with chiral nematic liquid crystal. The cell thickness gradually increases from left to right. The upper half represents the area without ITO, while the lower half represents the area with ITO.

**Supplementary note I: Details of numerical simulations and theoretical analysis**

We first simulated the static states of the skyrmions under DC field with different voltages. In experiment, the dimensionless electric field required to induce a uniform aligned background is $\tilde{E}_b = \sqrt{1.5}$, illustrated in Fig. 1. In simulations, when the electric field reaches $\tilde{E}_b = \sqrt{10/7}$, the background directors align vertically along with the direction of the electric field. Decreasing the voltages below $\tilde{E}_b$, the background directors start to twist under the effect of chirality. In both cases, the skyrmions are localized LC topological patterns embedded into the translational invariant, shown in Fig. 1c. Further decreasing the voltage, the skyrmions expand and finally convert to the cholesteric fingers, shown in Fig. S1.

Since skyrmions are localized topogical protection pattern, we assume that skyrmions have little effect on the bulk background and can be ignored. In addition, the period of AC field of 20 Hz is 50 ms and this corresponds approximately 15000 steps in simulations, so it is computational consuming to achieve the steady states of the whole skyrmions. To simplify the problem, we conducted simulations to measure the response of the background director field to the electric field in the absence of skyrmions. For the simulations of background director fields, we take account of the flexoelectric response and the dielectric response of LCs to the AC field. In the simulations, we applied the AC field vertically to the cell. When $\mod(t, \mathrm{T}) < \mathrm{T}/2$, the voltage $U = U_0 + U_b$; When $\mod(t, \mathrm{T}) \geq \mathrm{T}/2$, the voltage $U = -U_0 + U_b$. After the systems reach the steady states, we statistically collect the data of the systems. We observed that the director fields of the heliconical undergo a periodic oscillation during one period of AC fields. For better illustrations, we depicted director fields against time in Fig. 3. The bias dimensionless electric field of this case is $\tilde{E}_b = 0.54$.

We also measure the variations of the director fields by comparing the directors between adjacent frames, where $\delta\mathbf{n}(\mathbf{r},t) = \mathbf{n}(\mathbf{r},t+1) - \mathbf{n}(\mathbf{r},t)$. Here, the simulation time unit is dimensionless, so the minimum variation in time is 1 and time derivative can approximate to variation between two frames, $\partial_t \mathbf{n}(\mathbf{r},t) = \delta \mathbf{n}(\mathbf{r},t)$. In Fig. 1c, we found that the background directors at different height orientates to distinct directions. From Jonathan's theory, the velocity direction is perpendicular to both the direction of $\mathbf{n}_{BG}$ and the direction of $\partial_t \mathbf{n}_{BG}(\mathbf{r})$. Thus, the force acting on the 3D skyrmions are along distinct directions depending on the local background directors. To get the net motion of the skyrmions, we integral the force over the space. The vector $\mathbf{m}(t)$ is obtained from the summation of $\mathbf{n}(\mathbf{r},t) \times \partial_t \mathbf{n}(\mathbf{r},t)$ over space, $\mathbf{m}(t) = \Sigma_\mathbf{r} \mathbf{n}(\mathbf{r},t) \times \partial_t \mathbf{n}(\mathbf{r},t)$. The value of $m_x$ and $m_y$ against time is illustrated in Fig. SI. In this case, the integrals of $m_x$ and $m_y$ over one period are respectively -0.17 and 0.32, leading to the motion direction $\alpha = 96.6°$.

**Supplementary note II: Simulation unit analysis**

To nondimensionalization, we need characteristic parameters to represent length, time, electric charge. We take the simulation length unit as the nematic coherence length, $\xi_N = \sqrt{L_1/A}$. For traditional nematic materials, we use the following values, $A = 1.17 \times 10^5 \text{ J/m}^3$ and $L_1 = 4 \times 10^{-11} \text{ N}$ leading to $\xi_N = \sqrt{L_1/A} \approx 6.63 \text{ nm}$. In physical unit, rotational viscosity $\gamma_1$ of 5CB is $0.0777 \text{ Nm}^{-2}\text{s}$ and in simulation unit, the rotational viscosity is 0.1. These are matched by one iteration corresponding to $\Delta t \sim 3.5 \times 10^{-6} \text{s}$. For the unit for electric charge, we set the value of vacuum permittivity $\varepsilon_0$ in a simulation unit of 1, while in physical units, $35 \times \varepsilon_0 = 35 \times 8.85 \times 10^{-12} \text{C}^2\text{N}^{-1}\text{m}^{-2}$. We derive that 1 C in simulation units correspond to $0.98 \times 10^{-19} \text{ C}$ in physical units, equaling to $0.6125 \text{ e}$. Therefore, the flexoelectricity coefficient $e_{33} - e_{11}$ is 0.714 in simulation units matching with $11.35 \text{ pC/m}$ in physical units.

**Supplementary note III: Pitch measurement of chiral nematic liquid crystal**

For accurate determination of $p$, a "stripe–wedge" Grandjean–Cano cell is applied. Indium tin oxide (ITO) stripes with known thickness is coated on one of the plates of the wedge, Fig. S1(a). The thickness of the cell still varies continuously along the opening direction of the wedge but, importantly, discontinuously in the perpendicular direction, Fig. S1(b). The pitch can be determined through:

$$p = 2Ld/l \qquad (1)$$

Where $L$ is the distance between the closed disclinations, $d$ is the thickness of ITO surface and $l$ is the distance between the disclination in electrode region and non-electrode region.

# Description of Supplementary Movies

## Supplementary Movie 1.

Description: **Generation of skyrmions by gradually increasing the electric field.** $U_0 = 10.0 \text{ V}$, $f = 20 \text{ Hz}$, $U_b = 0$. The cell thickness is $6.7 \text{ μm}$. $T = 34°C$. Real time video rate.

## Supplementary Movie 2.

Description: **Generation of skyrmions from closed loops.** $U_0 = 10.0 \text{ V}$, $f = 20 \text{ Hz}$, $U_b = 0$. The cell thickness is $6.7 \text{ μm}$. $T = 34°C$. Real time video rate.

## Supplementary Movie 3.

Description: **Generation of skyrmions by abruptly changing the electric field.** $U_0 = 10.0 \text{ V}$, $f = 20 \text{ Hz}$, $U_b = 0$. The cell thickness is $6.7 \text{ μm}$. $T = 34°C$. Real time video rate.

## Supplementary Movie 4.

Description: **9 different motion directions of skyrmions under different bias voltage and the same square wave voltage and frequency.** $U_0 = 10.0 \text{ V}$, $f = 20 \text{ Hz}$. The cell thickness is $6.7 \text{ μm}$. $T = 34°C$. Real time video rate.

## Supplementary Movie 5.

Description: **Controlling the motion and cessation of skyrmions through abrupt changes in the electric field frequency from 20 Hz to 1 kHz under different bias voltage.** $U_0 = 10.0 \text{ V}$. The cell thickness is $6.7 \text{ μm}$. $T = 34°C$.

## Supplementary Movie 6.

Description: **Controlling the direction of skyrmions motion through abrupt**

**changes in the bias voltage.** $U_0 = 10.0 \text{ V}$. The cell thickness is $6.7 \text{ μm}$. $T = 34°C$.

## Supplementary Movie 7.

Description: **Trajectory of skyrmions in the way of "NJUPT" and "HKUST".** $U_0 = 10.0 \text{ V}$, $f = 20 \text{ Hz}$, $U_b = 0$.